# Thermochemistry of Alane Complexes for Hydrogen Storage: A Theoretical and Experimental Investigation

*Bryan M. Wong,*[1] *David Lacina,*[2] *Ida M. B. Nielsen,*[1] *Jason Graetz,*[2]* *and Mark D. Allendorf*[1]*

[1]Sandia National Laboratories, Livermore, CA 94551-0969, USA

[2]Brookhaven National Laboratory, Upton, NY 11973

*Corresponding author E-mail: mdallen@sandia.gov

**Abstract**

Knowledge of the relative stabilities of alane ($AlH_3$) complexes with electron donors is essential for identifying hydrogen storage materials for vehicular applications that can be regenerated by off-board methods; however, almost no thermodynamic data are available to make this assessment. To fill this gap, we employed the G4(MP2) method to determine heats of formation, entropies, and Gibbs free energies of formation for thirty-eight alane complexes with $NH_{3-n}R_n$ (R = Me, Et; $n$ = 0-3), pyridine, pyrazine, triethylenediamine (TEDA), quinuclidine, $OH_{2-n}R_n$ (R = Me, Et; $n$ = 0-2), dioxane, and tetrahydrofuran (THF). Monomer, bis, and selected dimer complex geometries were considered. Using these data, we computed the thermodynamics of the key formation and dehydrogenation reactions that would occur during hydrogen delivery and alane regeneration, from which trends in complex stability were identified. These predictions were tested by synthesizing six amine-alane complexes involving trimethylamine, triethylamine, dimethylethylamine, TEDA, quinuclidine, and hexamine, and obtaining upper limits of $\Delta G°$ for their formation from metallic aluminum. Combining these computational and experimental results, we establish a criterion for complex stability relevant to hydrogen storage that can be used to assess potential ligands prior to attempting synthesis of the alane complex. Based on this, we conclude that only a subset of the tertiary amine complexes considered and none of the ether complexes can be successfully formed by direct reaction with aluminum and regenerated in an alane-based hydrogen storage system.

# Introduction

Among the light metal hydrides, alane ($AlH_3$) is an attractive material for hydrogen storage, having a hydrogen weight percent of more than 10% and a decomposition temperature of ~100 °C.[1] In addition, $\alpha$-alane is kinetically stable in air for long periods of time.[2] However, regeneration of the hydride remains the major barrier to the use of this material. Specifically, rehydrogenation of the products of Reaction (1):

$$AlH_3 \rightarrow Al + 3/2\ H_2 \qquad (1)$$

is infeasible under practical conditions because of the extremely high equilibrium pressure of $H_2$ (~7 × 10³ atm at 298 K).[3] A promising solution to this problem was proposed by Graetz, et al., in which an intermediate reaction involving a complex between $AlH_3$ and an electron donor L, such as an amine or ether, is used to drive the uptake of hydrogen.[1,4] The resulting complex, a number of which are known,[5-10] would then be decomposed to regenerate $AlH_3$. An important requirement is that the complex must be sufficiently unstable that the Al–L bond(s) break before hydrogen is lost from $AlH_3$ itself. The first step in such a process was demonstrated by Graetz, et al. using triethylenediamine (TEDA) as the complexing agent and aluminum metal activated by a titanium catalyst:[4]

$$Al(Ti) + TEDA + 3/2\ H_2 \rightarrow Al(Ti)H_3\text{-}TEDA \qquad (2)$$

The alane-TEDA complex itself is unattractive as a hydrogen storage material, however. Its hydrogen storage capacity is low (2.6% by weight, assuming only aluminum-bound hydrogen is released), and its decomposition and regeneration kinetics are slow. A less stable adduct is therefore required that can be readily decomposed at temperatures below 100 °C without loss of hydrogen from the alane itself.

Although the structural chemistry of group 13 complexes has received considerable attention, very little quantitative information concerning the thermochemistry of these adducts is available. Other than the work of Graetz et al. just discussed,[4] there are only a few experimentally measured heats of formation.[11,12] Thus, one must rely primarily upon chemical intuition and empirical means to identify donor compounds that form complexes of appropriate stability for hydrogen storage. Further clouding the situation is the fact that multiple complex geometries are possible. Complexes with 1:1 (monomer) and 1:2 (bis) $AlH_3$:L structures are known, as well as bridged "dimer" structures (Figure 1). Given the large number of ethers and amines that could be used, identification of useful complexes by empirical means is a daunting task.

Computational chemistry offers an obvious alternative to extensive synthetic investigation. However, a new dilemma arises: many alane complexes of potential interest are large by the standards of molecular-orbital approaches, making them inaccessible to levels of theory needed to achieve chemical accuracy (±1 kcal mol$^{-1}$) in heats of formation and bond-dissociation energies. Previous theoretical work on alane complexes used either very high-level methods,[6,7,13-15] such as state-of-the art coupled-cluster calculations by Dixon et al.,[13-15] that cannot be routinely applied to the large molecules of interest here (for example, the complex $AlH_3$:(TEDA)$_2$ contains 44 atoms), or lower-level methods that do not provide sufficient accuracy.[16-18] To obtain valid comparisons among a range of these molecules, it is necessary to have accurate calculations both for reliable predictions of thermochemistry and calibration standards for benchmarking lower levels of theory.

In this paper we use high-level electronic structure calculations to obtain thermodynamic data for 13 alane-amine complexes, using the very accurate G4[19] and G4(MP2)[20] levels of theory, that allow us to assess their suitability for hydrogen storage. Structures, electronic energies, and temperature-dependent heats of formation, entropies, and Gibbs free energies for monomer, bis, and selected dimer complexes were obtained. Since these calculations are for gas-phase complexes, we then test the validity of the predicted trends by synthesizing and thermodynamically characterizing six amine-alane complexes that span the range of predicted stabilities. Using results of the ab intio calculations, we computed the thermodynamics of complexation for reactions (3) and (4) for the six cases considered experimentally:

$$AlH_3 + nL \rightarrow AlH_3:(L)_n; \quad n = 1, 2 \text{ (monomer or bis complex formation)} \quad (3a)$$

$$2AlH_3 + 2L \rightarrow LH_2Al(H_2)AlH_2L \text{ (dimer formation)} \quad (3b)$$

$$Al + 1.5\ H_2 + nL \rightarrow AlH_3:(L)_n; \quad n = 1, 2 \quad (4a)$$

$$2Al + 3H_2 + 2L \rightarrow LH_2Al(H_2)AlH_2L \quad (4b)$$

Reaction (3) corresponds stoichiometrically to solution-phase synthesis, in which the amine L is reacted with alane (absent solvent effects), while reaction (4) is defined for Al and $H_2$ in their standard states and corresponds to complex formation from the amine with solid aluminum. Thermodynamic data for these reactions provide a consistent way to evaluate complex stability independent of solvent effects and phase changes (e.g., precipitate formation), which are difficult to predict accurately. Together, the results give confidence in overall trends and enable criteria for selecting promising candidates for regenerable alane-based hydrogen storage to be determined.

In addition to this detailed treatment of alane-amine complexes, we also report data not available in the literature as an aid to future synthetic efforts. First, we include thermodynamic data for seven alane-ether complexes. For reasons discussed below, these complexes are not suitable for hydrogen storage and thus were not addressed experimentally. Second, we discuss trends in Al-H vibrational frequencies, which have been used to distinguish specific cluster geometries based on the value of the Al-H stretching frequencies. Finally, we briefly compare G4 predictions with the B3LYP density functional and the BAC-MP2 methods, two lower levels of theory often used to assess relative thermodynamic stability, and show that they do not fully capture the trends in complexation energies that occur with increasing alkyl substitution.

**Computational methods**

Heats of formation and complexation enthalpies were determined using the G4 and G4(MP2) composite techniques. The G4 model[19] combines high-level correlation/moderate basis set calculations with lower-level correlation/larger basis set calculations to approximate the results of a more expensive calculation. The composite G4 energy is obtained from results using CCSD(T), MP4, MP2, and HF calculations with progressively larger basis sets, and including first-order spin-orbit corrections for atoms and molecules, zero-point energy corrections (ZPE), and an empirical higher-level correction that depends on the number of paired and unpaired electrons. The G4(MP2) method[20] approximates full G4 theory by replacing the original MP4 calculations with MP2 theory, resulting in significant computational savings. G4 theory gives an average absolute deviation of 0.83 kcal/mol on the G3/05 experimental test set of energies.[19] The G4(MP2) method also performs well, with an average absolute deviation of 1.04 kcal/mol,[20] a level of accuracy that is actually better than the full G3 method. Thermal corrections were computed from standard formulas from statistical mechanics, and all computations were performed with the Gaussian09 program suite.[21] To benchmark less computationally demanding methods, we also calculated heats of formation and complexation enthalpies using the bond additivity correction (BAC) methods BAC-MP2[22] and BAC-MP4,[23-25] and with the B3LYP density functional method using a high-quality 6-311++G(2df,2dp) augmented triple-$\zeta$ basis set.[26,27] Optimized geometries and harmonic frequencies (corrected by a factor of 0.9854[28]) were determined at the B3LYP/6-311++G(2df,2dp) level of theory. Additional details are provided in the Supporting Information.

**Experimental methods**

**Synthesis and characterization of alane complexes.** Hydrogen was obtained from Praxair specified as 99.95% pure. The following were obtained from Sigma-Aldrich: THF (99.9% anhydrous); $AlCl_3$ (99.999%); *n*-undecane (99%); $LiAlH_4$ (reagent grade 95%); trimethylamine (TMA;99%); dimethylethylamine (DMEA; 99%);triethylamine (TEA; 99.5%); triethylenediamine (TEDA; 98%); hexamine (99%); and quinuclidine (97%). Titanium-catalyzed aluminum was prepared as described previously.[4]

Hydrogenation and dehydrogenation reactions were carried out in a 300 mL stainless steel stirred reactor (Parr Instruments) rated for 200 atm maximum operating pressure. Aluminum hydrogenation (reaction 4) occurs at room temperature and involves the direct reaction of $H_2$ with a slurry consisting of a solvent (100 ml of diethyl ether or tetrahydrofuran), a tertiary amine, and typically 1 g (0.037 mol) of catalyzed aluminum (2 mol% Ti) powder. $AlH_3$ complex yields were typically 40-60% based on the initial aluminum, with specific amine/solvent reactions producing higher yields. In all cases, the amount of hydrogen released during decomposition was equivalent to the amount of hydrogen uptake. The second hydrogenation cycle proceeded more quickly than the first and produced greater AlH$_3$ yields under similar conditions.

Fourier transform infrared spectroscopy (FTIR) using a Perkin-Elmer Spectrum One spectrometer, was performed to confirm the formation of the aluminum hydride adduct based on observation of the Al-H stretching modes (1650-1850 cm$^{-1}$). Powder XRD of solid alane complexes was performed using a Philips X-ray diffractometer with Cu Kα radiation.

***Trimethylamine alane (TMAA)*** was formed from 10.5 ml (0.119 mol) of liquid trimethylamine (TMA), diethyl ether, and an initial hydrogen pressure of 117.9 bar. Hydrogenation of the aluminum occurred over 24 hrs, reaching a final pressure of 104.6 bar ($\Delta P$ = -13.3 bar). FTIR spectra from both the solvated (in solution) and non-solvated solid products confirmed adduct formation with an Al-H stretch mode at 1705 cm$^{-1}$, as previously reported.[29,30] The decomposition of TMAA was performed at 373 K and was complete after ~ 2 hrs.

***Dimethylethylamine alane (DMEAA)*** was formed using 50 ml (0.461 mols) of liquid dimethylethlyamine (DMEA), diethyl ether, and an initial hydrogen pressure of 72.4 bar, using the method described previously.[31] The hydrogenation occurred over 56 hrs, reaching a final pressure of 65.3 bar ($\Delta P$ = -7.1 bar). FTIR of the liquid product confirmed the formation of the adduct with an Al-H stretch mode at 1710 cm$^{-1}$. Decomposition of DMEAA was complete after ~10 hrs at 295 K.

***Triethylamine alane (TEAA).*** We were unable to obtain thermodynamic data for this compound because it could not be formed via reaction (4). However, as reported previously,[31] TEAA can be synthesized by direct reaction of TEA with alane (reaction (3)) in various solvents. The FTIR spectrum confirmed adduct formation, with an Al-H stretch at 1777 cm$^{-1}$.

***Triethylenediamine alane (TEDAA)*** was formed using 17g (0.15 mol) of solid triethylenediamine (TEDA), THF, and an initial hydrogen pressure of 34.5 bar, using the method described previously.[4,21] The hydrogenation occurred over 80 hrs, reaching a final pressure of 24 bar ($\Delta P$ = -10.5 bar). XRD of the insoluble solid product confirmed the formation of TEDAA. Decomposition of TEDAA at 393 K was complete after 10 hrs.

*Hexamine alane (HexA)* was formed using 6g (0.043 mol) of hexamine (solid) in THF at a hydrogen pressure of 65.7 bar. Hydrogenation of the aluminum occurred over 48 hours, reaching a final pressure of 59 bar ($\Delta P$ = -6.7 bar). FTIR of the insoluble solid hexamine alane product confirmed adduct formation with an Al-H stretch mode at 1747 cm$^{-1}$. The decomposition of the hexamine alane adduct was performed at 353 K and was complete after ~ 3 hrs.

*Quinuclidine alane (quinA)* was formed using 5g (0.045 mol) of quinuclidine (solid) in THF at an initial hydrogen pressure of 64 bar. The hydrogenation reaction occurred over 24 hours, reaching a final pressure of 57.6 bar ($\Delta P$ = -6.4 bar). The solid product is soluble in THF and can be recovered by vacuum distillation. FTIR spectra of both the solvated and non-solvated products confirmed adduct formation with an Al-H stretch mode at 1700 cm$^{-1}$, similar to that previously reported.[11] Decomposition of quinA was complete after ~ 3hrs at 393 K.

**Predicted complex thermochemistry**

**Structures.** Optimized structures for selected monomer, bis, and dimer complexes computed at the G4(MP2) level of theory are illustrated in Figures 1 (TMAA), S1 (amine ligands), and S2 (ether ligands). The AlH$_3$ group is pyramidal in the monomer complexes, whereas the bis complexes have a trigonal bipyramidal configuration with a planar AlH$_3$ moiety in the center. The Al–N distances obtained in geometry optimizations at the G4(MP2) level of theory are listed in Table S1 for all of the complexes considered. For the monomeric complexes, the G4(MP2) Al–N bond lengths lie in the range 2.043-2.093 Å . For the bis complexes, these bonds are about 0.1 Å longer in most cases. Al–N bond distances in the symmetric bridged dimer complexes are comparable (within 0.05 Å) to their corresponding bis complexes. Within the AlH$_3$:NH$_{3-n}$R$_n$ series (R = Me, Et), each substitution of an R group for an H atom shortens the Al–N G4(MP2) bond distance, except for the last one, yielding AlH$_3$:NR$_3$, which causes a significant elongation of the bond due to steric hindrance. As will be seen below, this considerably destabilizes the TEAA complex relative to the other amine-alane complexes.

The G4(MP2) Al–O distances are also listed in Table S1, displaying distances in the 1.971-2.040 Å range for monomer complexes. Again, in the bis and bridged dimer complexes these bonds are in most cases about 0.1 Å longer than in the monomer complexes. Similarly, replacing an H atom with an alkyl group in the AlH$_3$:OH$_{2-n}$R$_n$ complexes decreases the Al–O bond length. These results are consistent with earlier computational investigations of AlH$_3$:NH$_3$,[32,33] AlH$_3$:N(CH$_3$)$_3$[6] and AlH$_3$:OH$_2$.[18]

**Heats of formation and trends in complex stability.** Heats of formation (298 K) computed at the G4-MP2 and G4 level of theory for monomer and dimer complexes considered here are given in Table 1 (complete thermodynamic data for all complexes and ligands considered here are given in Table S2), and values for selected dimer structures at the G4-MP2 level of theory are given in Table 2. There is good agreement between the measured heat of reaction for the first amine dissociation step of bis-TMAA reported by Heitsch (18.0 kcal mol$^{-1}$)[12] and the value computed from the G4(MP2) data (16.3 kcal mol$^{-1}$). We are unaware of any other measured gas-phase bond energies or heats of formation for comparison. However, the AlH$_3$NH$_3$ heats of formation predicted using other high-level methods (e.g. CCSD(T))[13-15,17] by Dixon et al. and others [6,32] are also in good agreement with the G4(MP2) values.

Using these data, $\Delta G°$, $\Delta H°$, and $\Delta S°$ of reaction (3) (Table 3) and reaction (4) (Table 4) were computed. Although $\Delta G°$ is the most complete indication of whether or not a reaction will occur, we first consider $\Delta H°(3)$, which is a useful guide to complex stability since $\Delta S°$ is relatively constant across all of the complexes and is the result whose accuracy is most dependent on the level of theory used. The values of $\Delta G°$ will become important when analyzing our experimental results (vide infra). Note that, for monomer formation, $\Delta H°$(3a) corresponds to the Al-L bond energy, while for bis complexes, ½($\Delta H°$(3a)) represents the average Al-L bond energy.

Several trends are evident from these results that are relevant to the synthesis of alane complexes for hydrogen storage purposes. First, the values of $\Delta H°$(3) and $\Delta G°$(3) are in all cases negative, indicating these complexes should be stable in spite of the relatively large decrease in entropy resulting from the change from two reactant molecules to one product molecule. Second, alane-amine complexes are significantly more stable than alane-ether complexes, regardless of the geometry of the complex. For example, $\Delta H°$ of reaction (3a) for monomer complexes ranges from -26 to -35 kcal mol$^{-1}$ for the amine complexes and from -17 to -26 kcal mol$^{-1}$ for the ether complexes. Third, in almost all cases, bis amine compounds are more stable than either monomer or dimer complexes on a per $AlH_3$ basis. This distinction disappears or is even reversed for ether complexes, suggesting that mixtures of all three geometries may form during synthesis. Fourth, dimer complexes differ very little in stability from monomer compounds; which of these geometries forms in practice may therefore depend on factors such as reaction kinetics or solvent. Finally, both $\Delta G°$(3) and $\Delta H°$(3) tend to become more negative with increasing alkyl substitution for both amines (primary < secondary < tertiary) and ether complexes (exceptions to this are discussed below). Complexes involving the largest amines (e.g. quinuclidine) and ethers (e.g. THF and dioxane) are the most stable of those considered. The thermodynamics of complexation indicate that $AlH_3:NH_3$ and $AlH_3:NEt_3$ are the least stable of the amines with respect to dissociation to $AlH_3$ and L. However, tertiary amines are desirable since 1,2-H$_2$ elimination from primary and secondary amines is predicted to be near thermoneutral,[11] leading to stable aluminum amine compounds (e.g. $Al_2NH_2$) instead of $AlH_3$. Thus, complexes with TEA and pyrazine, for which $\Delta H°$(3a) is 8–9 kcal/mol more positive than $AlH_3$-TEDA, should most easily allow $AlH_3$ regeneration, while complexes such as TEDAA and quinA will be the most difficult to regenerate.

The exceptions to these trends almost always occur when steric hindrance is a factor. In particular, tertiary amine complexes with alkyl substituents are either of comparable or lower stability than complexes with the corresponding primary and secondary amines. This effect is particularly evident in the case of TEAA ($AlH_3$:$NEt_3$), for which $\Delta G°$(3) is > 7 kcal/mol more positive than the corresponding value for $AlH_3$:$NH_2Et$ (Table 3). This large stability reduction is primarily due to the distorted geometry of the $NEt_3$ moiety within the complex compared to the unstrained geometry of the isolated $NEt_3$ molecule. More modest, but still significant, effects are evident in the series $AlH_3$:$NMe_n$ ($n = 0 - 3$) and for secondary ethers.

It is important to note that less demanding computational methods often used to predict main-group thermochemistry do not fully capture the trends in alane complex stability (Table S2). The agreement between BAC-MP4 heats of formation and the G4 values is poor compared with the G4(MP2) predictions, with differences as high as 7.6 kcal/mol ($AlH_3$:TEDA). The largest differences occur for the tertiary amine complexes $AlH_3$:$NMe_3$ and $AlH_3$:$NEt_3$ and the ether complexes $AlH_3$:$OMe_2$ and $AlH_3$:$OEt_2$. Moreover, whereas the G4 and G4(MP2) methods predict stabilization of the complexes upon substituting methyl groups for H (Figure 2), the BAC-MP4 method finds essentially no additional complex stabilization after addition of the first alkyl group. The popular B3LYP density functional method fares little better, even when the very large 6-311++G(2df,2pd) basis set is used. In this case, the predicted complexation enthalpies are all significantly more positive than the G4 values, and the predicted trends are somewhat different. In contrast, the maximum and average absolute differences between G4 and G4(MP2) complexation enthalpies are only 1.9 kcal/mol (AlH$_3$:(NMe$_3$)$_2$) and 1.4 kcal/mol, respectively. These comparisons justify the use of computationally intensive G4 methods when attempting to assess stability of these Lewis acid-based complexes.

## Alane complex synthesis, characterization, and thermodynamics

**Product identification based on Al-H stretching frequencies.** The assignment of alane complex geometries, particularly in solution, is often based on Al-H stretching frequencies, but definitive structural data are not always available to confirm these assignments. Consequently, we compared the gas-phase frequencies predicted by G4(MP2) with the vibrational spectra obtained from the alane-amine complexes synthesized by reaction (4) to provide evidence supporting both our assignments and previously reported ones in the literature. As seen in Table 5, the symmetric and asymmetric stretches of all complexes differ by < 20 cm$^{-1}$, which may not be observable in an FTIR or Raman spectrum at typical resolution, particularly if obtained in the condensed phase. Experimentally, Al-H frequencies in the 1750 – 1800 cm$^{-1}$ range are typically associated with monomer complexes, while complexes possessing the bis structure have lower Al-H frequencies in the 1700 – 1750 cm$^{-1}$ range, as do those of dimer complexes.[29,34] The G4(MP2) calculations confirm this. However, the predicted frequencies are systematically higher by 25 – 35 cm$^{-1}$, based on comparison with the measured gas-phase Al-H frequencies for monomer- and bis-TMAA[30,35] and monomer-DMEAA.[36,37] Predicted Al-H frequencies for monomer (1815 – 1837 cm$^{-1}$) and dimer (1821 – 1844 cm$^{-1}$) complexes are virtually the same, but the frequencies of bis complexes are significantly lower (60 – 90 cm$^{-1}$), between 1727 – 1777 cm$^{-1}$, suggesting this difference can be used as a structural diagnostic. Note, however, that the measured frequencies reported here are for the compounds in the solid or liquid (solvated) state, so some differences with respect to the gas-phase values are expected.

Turning to the amine complexes synthesized here, Al-H vibrational bands for both monomer and bis TMAA complexes are consistent with those expected for these geometries and confirm previous assignments.[29,30] The structure of the DMEA complex is less clear. Comparing the measured Al-H Raman frequencies (1725 – 1732 cm$^{-1}$) with the G4(MP2) results, a bis complex is suggested. However, we previously concluded that reaction (4) yields primarily the dimer, with a small amount of bis impurity.[31] This conclusion is based the instability of DMEAA at room temperature, which is consistent with dimer formation and may result from steric hindrance caused by the ethyl group. The measured TEAA frequencies are consistent with a monomer; as expected, the bis TEAA complex does not form due to steric hindrance.[11] The low Al-H frequencies in HexA and QuinA indicate bis complexes, in agreement with a previous report for QuinA.[11] We were unable to form the QuinA monomer by reaction (4); however, the IR and Raman data indicate that it does form via reaction (3a) (using a 1:1 mixture of solid AlH$_3$ with quinuclidine in solution). In this case, the measured frequencies agree well with those predicted by G4(MP2) (Table 5). Finally, the Al-H frequency of TEDAA appears to be an anomaly. The value we obtain, 1710 cm$^{-1}$, is consistent with a bis complex. However, in our previous work,[4] we obtained a product similar to that of Ashby,[5] who determined that a monomer was formed from a procedure similar to ours, but at a much higher pressure.

**Equilibrium pressure measurements**. The trends revealed by the G4(MP2) results suggest some of the complexes considered here could be useful for hydrogen storage. We therefore attempted to synthesize a cross-section of these complexes to first determine if they can be formed by direct hydrogenation (reaction 4) and second, whether the amine-alane can be separated from solvent so that AlH$_3$ can regenerated via the reverse of reaction (3) (denoted reaction (-3) in the following). Results of these experiments are summarized in Table 6, including estimates of $\Delta G°(4)$ at the reaction temperature. For DMEA, values of $\Delta G°(4)$ were determined for both the hydrogenation and dehydrogenation reactions, since these were performed at the same temperature (295 K). This allows an upper and lower bound on the free energy at that temperature to be determined from the equilibrium $H_2$ pressure using the following equation: $\Delta G°(4) = RT\ln P(H_2)$. For the TMA, hexamine, and quinuclidine reactions, only the upper bound on $\Delta G°_{295}(4)$ could be determined directly from hydrogen uptake measurements, as the dehydrogenation for these amines was performed at elevated temperatures (353 – 393 K). Lower bounds for these reactions could only be estimated and required an assumption for the value of $\Delta S°(4)$. Based on the value previously determined for the TEDA reaction,[4] we assumed a range of -27 – -31 cal $K^{-1}$ $mol(H_2)^{-1}$ for hydrogenation. Details of these calculations are provided in the Supporting Information.

**TMAA.** The value of $\Delta G°(4)$ is the upper bound only and was estimated from the pressure of $H_2$ at which uptake occurred to form the complex. It is likely that the actual value is much less than this number and is presumably negative, since the complex can be isolated and is stable at room temperature.

**DMEAA.** This complex decomposes at room temperature in the absence of solvent and an overpressure of $H_2$. Due to increased steric hindrance, the stability of this complex is intermediate between TMAA, which is stable at room temperature, and TEAA, which does not form by reaction (4). The upper and lower bounds for $\Delta G°(4)$ given in Table 6 were estimated at 298 K from the pressure of $H_2$ at which uptake occurred to form the complex. This value is valid only at this temperature because it was estimated from the point of decomposition. Although DMEAA is too unstable to be used for hydrogen storage directly, regeneration of AlH$_3$ was demonstrated via transamination to form the TEAA complex.[31]

**TEAA.** We were unable to obtain thermodynamic data for this compound because it is evidently too unstable to be formed via reaction (4).

**TEDAA.** As discussed above, thermodynamic data for the TEDAA complex were previously reported.[4] This complex is not usable for hydrogen storage because it is too stable; heating to regenerate the alane at 393 K results in hydrogen loss from AlH$_3$ itself.

**Hexamine alane (HexA).** The upper and lower bounds of $\Delta G°(4)$ were estimated from the pressure of H$_2$ at which uptake occurred to form the complex and the decomposition at 353 K; $\Delta G°(4)$ is valid only at this temperature, but we have made approximations to extrapolate to a value at room temperature.

**Quinuclidine alane (quinA).** The upper and lower bounds of $\Delta G°(4)$ were estimated from the H$_2$ uptake pressure during adduct formation (at room temperature) and the decomposition temperature of 393 K.

## Discussion

The measured $\Delta G°(4)$ values in Table 6 indicate the following order of stability for the six complexes produced via reaction (4) (note: the geometry and stoichiometry of the TEDAA complex were not determined experimentally in Ref. 4, but a DFT calculation reported there suggests that the monomer is stable in THF solution. The relative stability of bis and dimer complexes was not reported):

quinA (bis) > hexA (bis) > TEDAA (monomer?) > DMEAA (bis) > TMAA (bis) >> TEAA (monomer)

As expected from the G4(MP2) results, bis complexes are the predominant structures obtained experimentally. The overall order of stability predicted by G4(MP2) (numbers in parentheses are $\Delta G°(4)$ in kcal mol$^{-1}$) is:

hexA (bis, 0.4) > quinA (bis, 1.5) ~ TEDAA (bis, 1.7) > TMAA (monomer ~ bis, 3.2) ~ DMEAA (bis, 3.6) >> TEAA (monomer ~ dimer, 12.2)

These two series show that the general trend predicted by the calculated thermodynamic values is consistent with the experiments. Complexes with the largest amines (hexamine, quinuclidine, and TEDA) are significantly more stable than those with ethyl substituents (DMEAA and TMAA), which is reflected in the equilibrium constants $K_{eq}(4)$ that can be computed from the data in Table 4. For example, the somewhat higher stability of TEDAA (smaller $\Delta G°(4)$) relative to TMAA leads to $K_{eq}(4)$ for TEDAA that is a factor of 13 larger at 298 K. In contrast, the much lower stability of TEAA (more positive $\Delta G°$) relative to DMEAA results in $K_{eq}(4)$ that is ~10$^6$ smaller for TEAA, indicating TEAA formation is very unfavorable. Such conclusions must be qualified by noting that factors other than thermodynamics may also play a role in complex formation. In particular, the possible formation of the TEDAA monomer, although not the thermodynamically favored product based on the gas-phase $\Delta G°$, could well be driven by the immediate precipitation of the solid. In contrast, the quinA complex remains in solution until the solvent is removed by evaporation, which could provide sufficient time for the bis complex to form.

The extent to which the six amine complexes can be decomposed to regenerate alane via reaction (-3) is obviously influenced by $\Delta G°$ of the reaction, but here it appears that the reaction kinetics are also important. The bis-DMEAA, mono-TEDAA, and bis-hexA complexes illustrate this point. Of the three, bis-DMEAA decomposes spontaneously at room temperature and goes to completion within 10 hours, consistent with its lower $\Delta G°(3)$ (-27.1 kcal mol$^{-1}$) relative to bis-hexA ($\Delta G°(3)$ = -31.8 kcal mol$^{-1}$). This complex is the most stable of these three and requires heating to 353 K to decompose fully within 3 hours. TEDAA on the other hand, requires heating to 120 °C for 10 hours to completely decompose. Monomer-TEDAA is not only less stable ($\Delta G°(3)$ = -24.2 kcal mol$^{-1}$) than either DMEAA or hexA, but is also significantly less stable than its bis form ($\Delta G°(3)$ = -29.8 kcal mol$^{-1}$). These data therefore suggest that it is actually bis-TEDAA that forms via reaction (4) in the titanium-catalyzed synthesis reported in Ref. 4 (which used excess TEDA), consistent with the IR frequencies (see above). Previously, it was speculated that the high melting point of TEDA might be the cause of the slow decomposition kinetics of the TEDAA adduct, since H$_2$ would have to diffuse through solid product TEDA to escape.[4]

Based on the results above, we can establish a criterion for selecting alane complexes appropriate for hydrogen storage. A balance must be struck between values of $\Delta G°(-3)$ and $\Delta G°(4)$ such that the complex is sufficiently stable to form via reaction (4), but not so stable that it cannot be decomposed by reaction (-3). One way to express this relationship is the correlation between $\Delta G°(-3)$ and $\Delta G°(4)$ shown in Figure 3. Here, the G4(MP2) data fall on a straight line, with the y-axis intercept equivalent to $(2/3)\Delta G°_f(AlH_3)$ and a slope of -0.67, resulting from the fact that the sum of reactions (-3) and (4) is Al(s) + (3/2)H$_2$ $\rightarrow$ AlH$_3$(g). Quantitative agreement between predicted and measured $\Delta G°$ is neither expected nor achieved in all cases, based upon the various factors discussed above that affect the thermodynamics and kinetics of these reactions. Nevertheless, even the fact that a given complex forms *at all* via reaction (4) can be used to establish a correlation between reaction thermodynamics and potential utility for hydrogen storage.

The correlation in Figure 3 indicates that there is a region in which both $\Delta G°(-3)$ and $\Delta G°(4)$ are appropriate for hydrogen storage applications, the boundary of which is roughly demarcated by the vertical dashed line in the figure. Empirically, we find that amine complexes within this region can be formed via reaction (4) and decomposed at modest temperatures ($\leq$ 100 °C) and reaction times ($\leq$ 3 hours). Complexes to the left of the line are either too unstable to form (high value of $\Delta G°(4)$) or decompose too quickly (low value of $\Delta G°(-3)$) to be useful. Alternatively, complexes to the right of the line may be too stable, requiring high temperatures to decompose that lead to hydrogen loss from $AlH_3$. This correlation is borne out by experiment: the TEAA complexes (arrows in Fig. 3), which have the highest predicted $\Delta G°(4)$, do not form at all. Similarly, we expect ether complexes are too unstable to form by reaction (4), as they lie well to the left of the line. TEDAA is an exception; the monomer complex can be formed by reaction (4), but its decomposition kinetics are too sluggish for it to be useful in hydrogen storage. The best case examined to date involves the transamination reaction between TMAA and DMEA, which allows DMEAA to form while the volatile TMA is removed under vacuum. These two complexes lie on the border separating the two regions. Since all of the amine complexes in Table 1 that we did not synthesize have $\Delta G°(-3) < 27$ kcal mol$^{-1}$, it is unlikely that any of these will form via reaction (4). We conclude that, to improve upon the TMAA/DMEA cycle, tertiary amines that are either more prone to precipitate than DMEAA (suggesting larger molecules) or are somewhat less stable (suggesting higher steric hindrance) are possibilities.

Examples include dimethylpropylamine and diethylmethylamine. An additional possibility is to find reaction conditions that produce the bis-TEDAA, since this complex falls to the right of the line.

We note that, in comparing the G4(MP2) predictions with our experimental results, we cannot rule out the possibility that more than one alane complex forms during synthesis. Indeed, we speculated previously that the reaction of DMEA with $AlH_3$ could produce a mixture of bis and dimer complexes.[31] If comparable amounts of two different products were formed by any of the reactions examined here, we would expect that the measured thermodynamics would not necessarily reflect those of an individual reaction. However, there is no clear evidence from vibrational spectroscopy that this occurs. In any case, the intent of the calculations is to approximate the actual reaction using relatively simple model systems so that trends in stability can be predicted. The results discussed above suggest that the predicted trends are consistent with experimental results.

**5. Concluding Remarks**

Using high-level electronic structure methods, we determined heats of formation and complexation thermodynamic data for alane complexes with amines and ethers, allowing their potential for hydrogen storage applications to be evaluated. Monomer, bis, and dimer complex geometries were considered and $\Delta G°$ computed for the formation and decomposition reactions relevant to hydrogen storage applications. We also synthesized five amine-alane complexes and obtained upper limits of $\Delta G°(4)$. These high-level thermochemical data, which were previously unavailable for all but the TMAA complex, provide useful guidance for synthetic efforts and a rational direction towards identifying alane complexes with the desired properties for hydrogen storage purposes. In particular, they enable trends in complex stability to be determined. Predicted vibrational frequencies also demonstrate the extent to which Al-H stretching frequencies can be used to distinguish bis complexes from either monomer or dimer structures. Our results can be used to rationalize several experimental observations, including the inability to form TEAA and the feasibility of the TMAA/DMEA transamination cycle. Finally, we determined a criterion for assessing potential ligands prior to attempting synthesis of the alane complex. Although the G4(MP2) is computationally a relatively expensive method, it was successfully applied to a wide range of amines and ethers. Consequently, it should be feasible to extend such calculations to other ligands, thereby reducing the extent of synthetic efforts required to identify an optimal alane-based hydrogen storage system.

**Acknowledgements.** This work was supported by the U.S. Department of Energy Fuel Cell Technologies Program through the Sandia Metal Hydride Center of Excellence. Sandia is a multiprogram laboratory operated by Sandia Corporation, a Lockheed Martin Company, for the United States Department of Energy's National Nuclear Security Administration under Contract

DE-AC04-94AL85000. D.L. and J.G. acknowledge support from the Office of Basic Energy Sciences, U.S. Department of Energy under Contract No. DE-AC02-98CH1-886.

**Supporting Information Available:** G4(MP2) geometries and zero-point energies for isolated ligands and complexes; optimized Al–N and Al–O bond distances for alane complexes at the HF/6-31G* and G4(MP2) levels of theory; heats of formation ($\Delta H_f$) at 0 and 298.15 K for isolated ligands computed at the G4(MP2) level of theory; complexation enthalpies at 0, 298.15, 325, 350, 375, and 400 K for monomer and bis complexes computed at the G4(MP2) level of theory; BAC-MP4 complexation enthalpies; G4(MP2) total energies; G4(MP2) geometries and zero-point energies for isolated ligands; calculation of the lower limit of $\Delta G°(4)$. This material is available free of charge via the Internet at http://pubs.acs.org.

**TABLE 1**: Heats of formation ($\Delta H°_f$) at 298.15 K for Al:N complexes (upper table section) and Al:O complexes (lower table section) computed at the G4 and G4(MP2) levels of theory.

| | Monomer | | Bis Complex | |
|---|---|---|---|---|
| | G4 | G4(MP2) | G4 | G4(MP2) |
| Ligand | $\Delta H°_{f,298.15}$ (kcal/mol) | $\Delta H°_{f,298.15}$ (kcal/mol) | $\Delta H°_{f,298.15}$ (kcal/mol) | $\Delta H°_{f,298.15}$ (kcal/mol) |
| NH$_3$ | -7.6 | -6.1 | -28.5 | -26.6 |
| NH$_2$Me | -5.8 | -4.0 | -24.5 | -21.9 |
| NHMe$_2$ | -7.0 | -4.9 | -26.7 | -23.7 |
| NMe$_3$ | -10.2 | -7.9 | -33.3 | -29.8 |
| NH$_2$Et | -13.0 | -11.1 | -36.4 | -33.9 |
| NHEt$_2$ | -19.2 | -17.1 | —[a] | -46.3 |
| NEt$_3$ | -20.6 | -18.1 | —[a] | -49.8 |
| Pyridine | 32.4 | 33.7 | 53.3 | 54.4 |
| Pyrazine | 51.2 | 52.8 | 87.2 | 89.1 |
| TEDA | 17.6 | 20.3 | —[a] | 26.7 |
| Quinuclidine | -6.9 | -4.6 | —[a] | -21.9 |
| OH$_2$ | -45.7 | -44.4 | -111.1 | -109.4 |
| OHMe | -40.5 | -38.7 | -98.7 | -96.0 |
| OMe$_2$ | -38.2 | -36.1 | -93.4 | -90.2 |
| OHEt | -49.1 | -47.2 | -115.0 | -112.2 |
| OEt$_2$ | -53.7 | -51.4 | -112.4 | -118.8 |
| OMeEt | -45.8 | -43.6 | -106.8 | -103.5 |
| Dioxane | -70.4 | -67.8 | -158.3 | -154.1 |
| THF | -40.0 | -37.9 | -94.8 | -91.5 |

[a] Computation was not feasible

**TABLE 2**: Heats of formation ($\Delta H°_f$) at 298.15 K for selected dimer complexes at the G4(MP2) level of theory.

| Ligand in Dimer Complex | G4(MP2) $\Delta H°_{f,298.15}$ (kcal/mol) |
|---|---|
| NMe$_3$ | -25.2 |
| NEt$_3$ | -45.7 |
| DMEA | -34.9 |
| TEDA | 31.2 |
| Quinuclidine | -17.8 |
| Hexamine | 79.0 |
| OMe$_2$ | -85.5 |
| OEt$_2$ | -114.5 |
| THF | -87.6 |

**TABLE 3**: Reaction (3) enthalpies ($\Delta H°$), entropies ($\Delta S°$), and free energies ($\Delta G°$) at 298.15 K for Al:N complexes (upper table section) and Al:O complexes (lower table section) computed at the G4(MP2) level of theory.

| Ligand | Monomer Formation | | | Bis Complex Formation | | | Dimer Formation | | |
| --- | --- | --- | --- | --- | --- | --- | --- | --- | --- |
| | $\Delta H°_{298.15}$ (kcal/mol) | $\Delta S°_{298.15}$ (cal/(mol·K)) | $\Delta G°_{298.15}$ (kcal/mol) | $\Delta H°_{298.15}$ (kcal/mol) | $\Delta S°_{298.15}$ (cal/(mol·K)) | $\Delta G°_{298.15}$ (kcal/mol) | $\Delta H°_{298.15}$ (kcal/mol) | $\Delta S°_{298.15}$ (cal/(mol·K)) | $\Delta G°_{298.15}$ (kcal/mol) |
| $NH_3$ | -27.1 | -30.9 | -17.9 | -37.4 | -62.3 | -18.8 | | | |
| $NH_2Me$ | -30.8 | -32.7 | -21.0 | -44.2 | -69.8 | -23.4 | | | |
| $NHMe_2$ | -32.8 | -33.9 | -22.6 | -48.1 | -73.2 | -26.3 | | | |
| $NMe_3$ | -33.6 | -36.0 | -22.9 | -49.9 | -75.0 | -27.6 | -76.7 | -105.5 | -45.2 |
| $NH_2Et$ | -31.2 | -32.5 | -21.5 | -42.9 | -67.8 | -22.7 | | | |
| $NHEt_2$ | -32.5 | -34.8 | -22.1 | -45.9 | -73.4 | -24.0 | | | |
| $NMe_2Et$ | -33.9 | -36.9 | -22.9 | -50.5 | -78.4 | -27.1 | -77.3 | -111.4 | -44.1 |
| $NEt_3$ | -26.7 | -41.5 | -14.3 | -35.7 | -83.8 | -10.7 | -62.9 | -115.9 | -28.3 |
| Pyridine | -30.3 | -30.1 | -21.3 | -42.3 | -65.7 | -22.7 | | | |
| Pyrazine | -27.1 | -28.4 | -18.6 | -39.3 | -62.2 | -21.0 | | | |
| TEDA | -34.8 | -35.7 | -24.2 | -52.2 | -75.2 | -29.8 | -79.0 | -106.4 | -47.2 |
| Quinuclidine | -35.6 | -32.4 | -25.9 | -52.6 | -75.0 | -30.2 | -79.8 | -105.2 | -48.4 |
| Hexamine | -33.8 | -31.9 | -24.2 | -52.1 | -67.9 | -31.8 | -78.0 | -103.9 | -47.0 |
| $OH_2$ | -18.1 | -25.7 | -10.5 | -25.6 | -60.5 | -7.5 | | | |
| OHMe | -22.1 | -30.2 | -13.1 | -31.7 | -66.2 | -11.9 | | | |
| $OMe_2$ | -23.7 | -30.2 | -14.7 | -34.0 | -65.6 | -14.5 | -60.6 | -97.6 | -31.5 |
| OHEt | -23.0 | -30.3 | -14.0 | -19.2 | -21.6 | -12.7 | | | |
| $OEt_2$ | -24.3 | -32.2 | -14.7 | -33.3 | -71.0 | -12.1 | -60.3 | -102.4 | -30.0 |
| OMeEt | -24.6 | -30.8 | -15.4 | -34.2 | -66.5 | -14.3 | | | |
| Dioxane | -24.2 | -30.4 | -15.1 | -35.4 | -67.2 | -15.4 | | | |
| THF | -26.2 | -30.7 | -17.1 | -37.0 | -66.6 | -17.1 | -64.3 | -101.6 | -34.0 |

**Table 4.** Predicted G4(MP2) thermodynamic data (298.15 K) for the formation of alane-amine complexes according to reaction (4).

|  | Monomer | | | Bis | | | Dimer | | |
| --- | --- | --- | --- | --- | --- | --- | --- | --- | --- |
|  | $\Delta H^{\circ a}$ | $\Delta S^{\circ b}$ | $\Delta G^{\circ a}$ | $\Delta H^{\circ a}$ | $\Delta S^{\circ b}$ | $\Delta G^{\circ a}$ | $\Delta H^{\circ a}$ | $\Delta S^{\circ b}$ | $\Delta G^{\circ a}$ |
| TMAA | -1.5 | -26.6 | 6.4 | -12.4 | -52.6 | 3.2 | -4.7 | -37.7 | 6.6 |
| DMEAA | -1.8 | -27.2 | 6.3 | -12.8 | -54.9 | 3.6 | -4.9 | -39.7 | 6.9 |
| TEAA | 3.1 | -30.3 | 12.1 | -2.9 | -58.4 | 14.5 | -0.1 | -41.2 | 12.2 |
| TEDAA | -2.4 | -26.4 | 5.5 | -14.0 | -52.7 | 1.7 | -5.5 | -38.1 | 5.9 |
| HexA | -1.7 | -23.9 | 5.5 | -13.9 | -47.8 | 0.4 | -5.2 | -37.2 | 5.9 |
| QuinA | -2.9 | -24.2 | 4.4 | -14.2 | -52.6 | 1.5 | -5.7 | -37.7 | 5.5 |

[a] kcal mol($H_2$)$^{-1}$. [b] cal mol($H_2$)$^{-1}$ K$^{-1}$.

**Table 5.** Terminal Al–H stretching frequencies (cm$^{-1}$) computed at the G4(MP2) level of theory, corrected by a factor of 0.9854,[28] and corresponding measured frequencies in the solid state.

| Complex | Monomer complex | | Bis complex | | Dimer complex | | Experimental frequencies (IR and Raman) |
|---|---|---|---|---|---|---|---|
| | Sym | Asym | Sym | Asym | Sym | Asym | |
| **TMAA** | 1837 | 1822 | 1750 | 1738 | 1837 | 1830 | Monomer: 1795;[b] bis: 1705[b] |
| **TEAA** | 1836 | 1820 | 1777 | 1767 | 1844 | 1838 | 1801 (1767)[a]; 1777[b] |
| **DMEAA** | 1836 | 1820 | 1755 | 1742 | 1838 | 1832 | 1710[b]; 1725-1732[a] |
| **TEDAA** | 1834 | 1817 | 1744 | 1732 | 1831 | 1824 | 1712;[a] 1710[b] |
| **Hexamine alane** | 1836 | 1819 | 1743 | 1730 | 1834 | 1827 | 1747[b] |
| **Quinuclidine alane** | 1832 | 1815 | 1740 | 1727 | 1829 | 1821 | Monomer: 1792[a], 1765[b] bis: 1700[b] |

[a] Raman. [b] IR.

**Table 6.** Experimental thermodynamic data for the formation of alane-amine complexes according to reaction (4).

| Complex | Geometry | Product phase | $P_{uptake}$ (bar) | $T_d$ (K)[a] | $\Delta G°_{295}(4)$[b] |
|---|---|---|---|---|---|
| TMAA | Bis complex | Solid in diethyl ether | 104.6 | 295 | $\leq 2.7$ |
| DMEAA | Bis complex | Liquid in THF | 57.3 | 295 | 1.4 – 2.4 |
| TEAA | Monomer | Liquid in TEA | -- | -- | --[d] |
| TEDAA | Monomer (?)[e] | Insoluble solid in THF | 24 | 393 | -1.1 |
| HexA | Bis complex | Insoluble solid in THF | 59 | 353 | -0.5 – 2.4 (-0.5 – 0)[c] |
| QuinA | Bis complex | Solvated solid in THF | 57.6 | 393 | -1.5 – 2.4 (-1.5 – 0)[c] |

[a] Decomposition temperature of the complex. [b] Data are given in kcal per mol $H_2$. [c] Complex is stable at room temperature, so $\Delta G°(4)$ is presumed to be negative at 295 K. [d] No reaction, so value could not be determined. [e] The geometry and stoichiometry of the TEDAA complex were not determined experimentally in Ref. 4; see text.

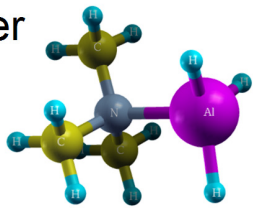

monomer

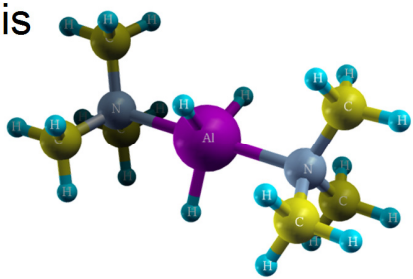

bis

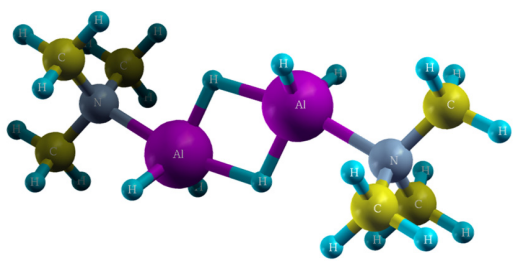

dimer

**Figure 1.** Complex geometries, using TMAA as an example.

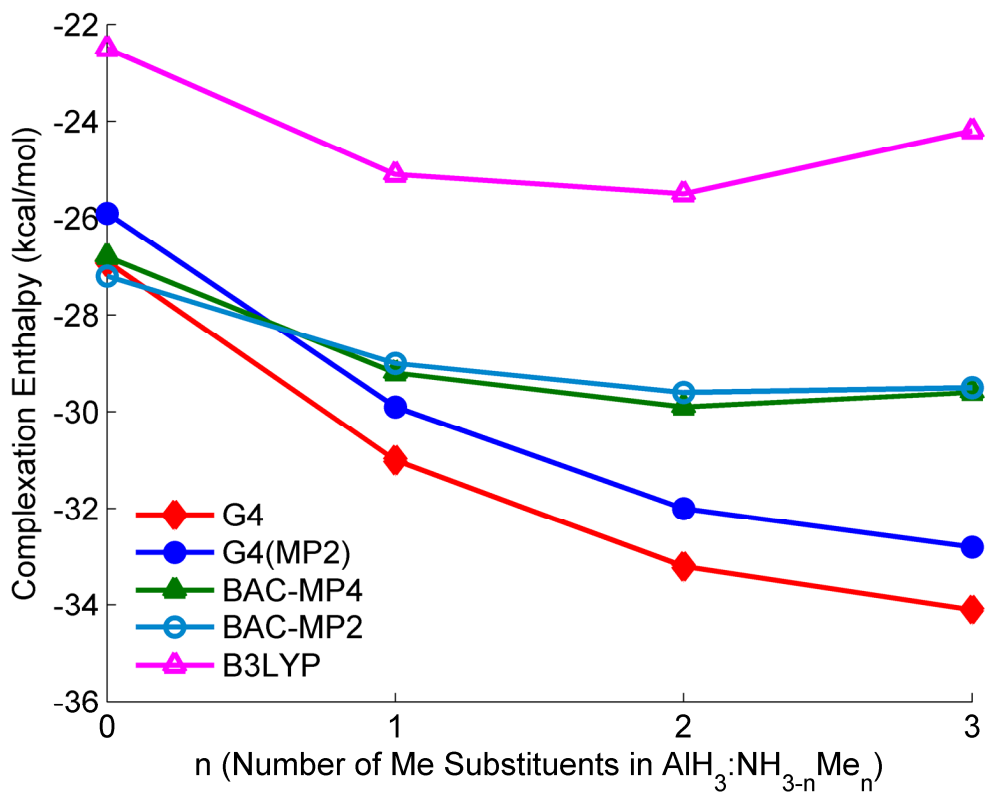

**Figure 2.** Complexation enthalpies as a function of *n*, the number of Me substituents in AlH$_3$:NH$_{3-n}$Me$_n$. Among all the methods considered, only the G4 and G4(MP2) methods predict a monotonic stabilization as a function of *n*.

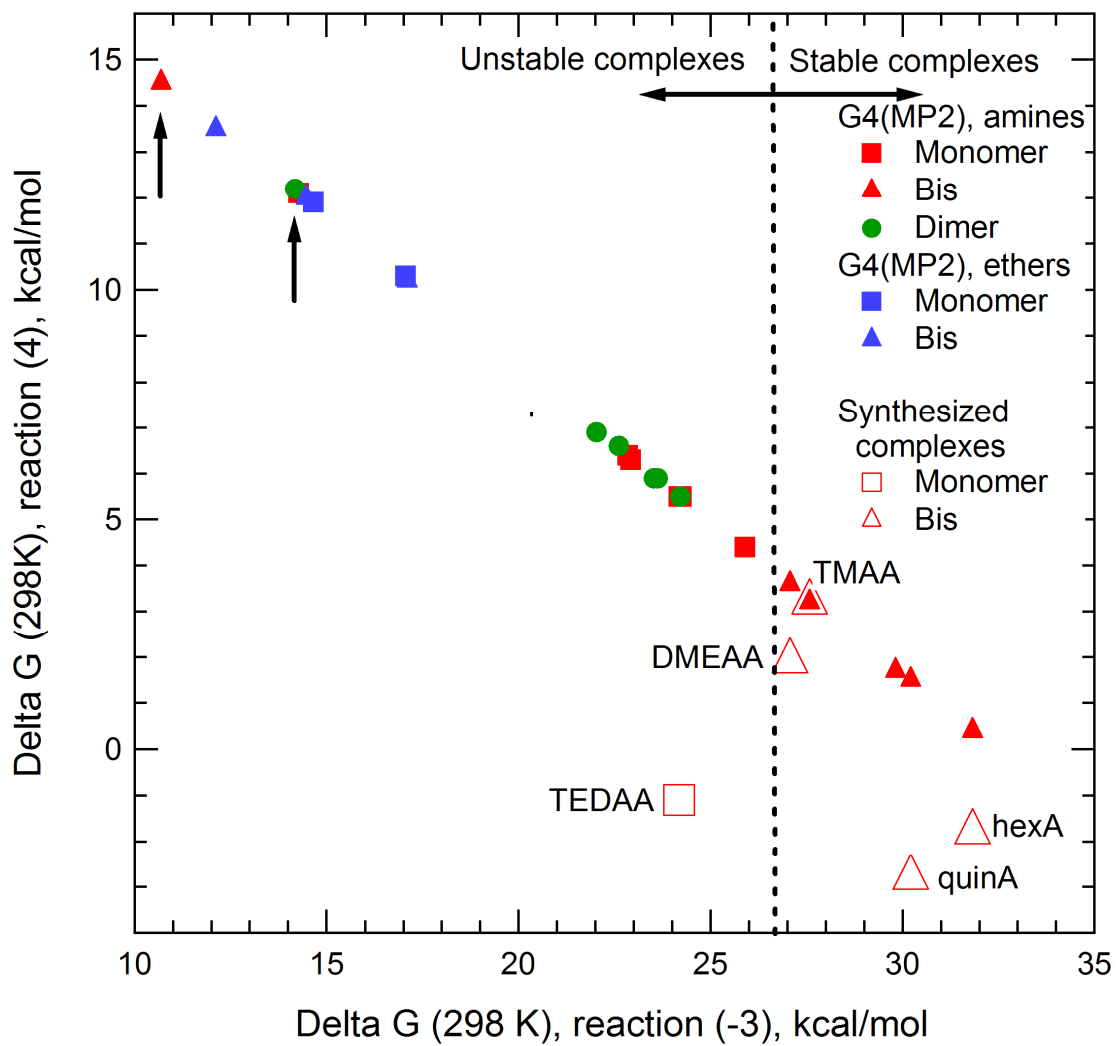

**Figure 3.** Correlation between G4(MP2) $\Delta G°(-3)$ and $\Delta G°(4)$, with experimental values given in Table 6. Labels on the plot (e.g. DMEAA) correspond to experimental values indicated by the blue open symbols. Vertical arrows indicate values for the monomer (left) and bis (right) TEAA complexes.